\RequirePackage[2020-02-02]{latexrelease}
\documentclass{iau}

\usepackage{amsmath}
\usepackage{graphicx}
\usepackage{multirow}
\usepackage{floatrow}

\usepackage{gensymb}
\usepackage{enumerate}

\newcommand{\kms}{\ifmmode{\,\rm{km}\, \rm{s}^{-1}}\else{$\,$km$\,$s$^{-1}$}\fi}
\newcommand{\msun}{\ifmmode{~\rm{M}_{\odot}}\else{M$_{\odot}$}\fi}
\newcommand{\mstar}{\ifmmode{M_{\star}}\else{$M_{\star}$}\fi}

\newcommand{\gaia}{\textit{Gaia}}

\newcommand{\monh}{[{\rm M}/{\rm H}]}
\newcommand{\afe}{[\alpha/{\rm Fe}]}
\newcommand{\feh}{[{\rm Fe}/{\rm H}]}

\newcommand{\mgfe}{[{\rm Mg}/{\rm Fe}]}

\newcommand{\yocc}{y_{\rm O}^{\rm cc}}

\newcommand{\Modot}{\dot{M}_{\rm O}}

\newcommand{\mdotstar}{\dot{M}_*}

\newcommand{\tauIa}{\tau_{\rm Ia}}
\newcommand{\Zo}{Z_{\rm O}}

\newcommand{\Zoeq}{Z_{\rm O,eq}}
\newcommand{\Zosun}{Z_{{\rm O},\odot}}

\newcommand{\taustar}{\tau_*}

\newcommand{\tausfh}{\tau_{\rm sfh}}

\newcommand{\Gyr}{\,{\rm Gyr}}

\newcommand\aap{A\&A}                
\newcommand\aj{AJ}                   
\newcommand\apj{ApJ}                 
\newcommand\apjl{ApJ}                
\newcommand\apjs{ApJS}               
\newcommand\araa{ARA\&A}             
\newcommand\fcp{Fundamentals Cosmic Phys.}  
\newcommand\mnras{MNRAS}             

\begin{document}

\lefttitle{Weinberg}
\righttitle{Proceedings of the International Astronomical Union: \LaTeX\ Guidelines for~authors}

\jnlPage{1}{7}
\jnlDoiYr{2021}
\doival{10.1017/xxxxx}

\aopheadtitle{Proceedings IAU Symposium}
\editors{F. Tabatabaei,  B. Barbuy \&  Y. Ting, eds.}

\title{New Dimensions of Galactic Chemical Evolution}

\author{David H. Weinberg$^{1}$}
\affiliation{$^1$Department of Astronomy and CCAPP, Ohio State University, \\
140 W. 18th Ave., Columbus, OH 43210, USA.  email: {\tt weinberg.21@osu.edu}\\}



\begin{abstract}
Dramatic recent progress in understanding galactic chemical evolution (GCE) has
been driven partly by direct observations of the distant past with 
{\textit{HST}} and {\textit{JWST}} and partly by archeaological interpretation
of stellar abundances from giant high-resolution spectroscopic surveys (APOGEE, GALAH) and
the complementary power of \gaia\ astrometry and photometry.  Focusing on archaeology, I
give a rapid-fire, and I hope synthesizing, review of work my collaborators and I
have done on theoretical modeling and observational interpretation.  I discuss
(1) the interleaved but distinguishable roles of stellar scale astrophysics and
galactic scale astrophysics in governing GCE, (2) the use of abundance ratio trends to empirically infer nucleosynthetic yields, (3) the uncertainty in the overall
scale of yields and its degeneracy with the importance of galactic outflows,
(4) the emergence of equilibrium in GCE, (5) the dimensionality of the stellar 
distribution in chemical abundance space, and (6) insights from chemical abundances
on the early history of the Milky Way, including measurements of the intrinsic
scatter of abundance ratios in metal-poor stars ($-2 \leq \feh \leq -1$) 
suggesting that a typical halo star at this metallicity is enriched by the
products of $N\sim 50$ supernovae mixed over $\sim 10^5 M_\odot$ of
star-forming gas.
\end{abstract}


\maketitle

\section{Introduction}

\noindent 

I am a relative latecomer to the field of galactic chemical evolution (GCE),
after working on large scale structure and other aspects of galaxy formation
through most of my career.  This late arrival has advantages and disadvantages.
The principal advantage is fresh perspective informed by my experience in other
fields.  The principal disadvantage is my sometimes limited knowledge of what
has come before, which in a voluminous field like GCE is a lot
(e.g., Tinsley 1980, Prantzos \& Aubert 1995, Pagel 1997, Kobayashi et al. 2006, Matteucci 2012).  
Fortunately, I have had wonderful
collaborators to help mitigate my ignorance.  Even more fortunately,
dramatic advances in GCE data sets have created radically new opportunities
for modeling and interpretation.  For me personally, the APOGEE survey
(Majewski et al. 2017) of SDSS-III (Eisenstein et al. 2011), 
SDSS-IV (Blanton et al. 2017), and SDSS-V (Kollmeier et al. 2017) has
been most important, but similar advances are coming from 
GALAH (Buder et al. 2021), \gaia\ (Gaia Collaboration 2023),
SEGUE (Yanny et al. 2013, Rockosi et al. 2023), 
LAMOST (Luo et al. 2015), H3 (Conroy et al. 2019),
and numerous observing programs that target specific stellar
populations in the Milky Way and its neighbors.

It is a long time since I wrote a conference proceedings, but IAU Symposia have
always been the gold standard of these, and this is my first opportunity to
contribute to one!  As in my talk, I will use this opportunity to run
lightly over a lot of ground, and to make editorial comments I might never
get past a stern journal referee.

\section{The Structure of the GCE Problem}
\label{sec:structure}

\noindent 
A GCE model relies on ``stellar scale'' astrophysics that determines the 
IMF-averaged yields and delay time distribution (DTD) of different enrichment
channels, and on ``galactic scale'' astrophysics such as accretion and star
formation history, star formation laws, outflows, radial gas flows, and
redistribution of stars by radial migration, disk heating, and mergers.
For compactness, I will refer to these components as SSA and GSA,
respectively.  Distinct GCE models may differ in their SSA assumptions,
their GSA assumptions, or both.  Model predictions can be tested against a 
wide range of observables in the Milky Way and other galaxies, including
abundance ratio trends (e.g., $\afe$ vs. $\feh$), distribution functions
(e.g., $p(\feh)$, $p(\afe)$), and the dependence of these on a stellar population's
age, galactic position, and kinematics.

\begin{table}
    \centering
    \begin{tabular}{lr}
    Abbreviation & Meaning \\
    \hline
    AGB & asymptotic giant branch \\
    CCSN & core collapse supernova(e) \\
    DTD & delay time distribution \\
    GCE & galactic chemical evolution \\
    GSA & galactic scale astrophysics \\
    IMF & initial mass function \\
    ISM & interstellar medium \\
    SFE & star formation efficiency \\
    SFH & star formation history \\
    SFR & star formation rate \\
    SNIa & Type Ia supernova (e) \\
    SSA & stellar scale astrophysics \\
    \hline
    \end{tabular}
    \caption{A partial list of abbreviations used}
    \label{tbl:acronyms}
\end{table}

This structure is reminiscent (to this reminiscer, anyway) of galaxy formation
in the late '80s and early '90s.  Models depended
both on poorly known cosmological inputs --- such as $\Omega_m$, $\Omega_b$,
the primordial power spectrum, and the nature of dark matter --- and on galactic
scale modeling of gas cooling, angular momentum, feedback, outflows, mergers,
and so forth.  They could be tested against observations of galaxy clustering
and galaxy properties over a range of redshifts, and the interesting challenge
was to learn about {\it both} cosmology and galaxy scale physics despite
the tradeoffs between the two.  

If nucleosythetic yields were a solved theoretical problem, then our GCE models
could take them as robust inputs and focus on constraining GSA --- rather
like galaxy formation today, where the uncertainties in cosmological parameters
are mostly negligible compared to the uncertainties in ``gastrophysics.''
However, the physics challenges of nucleosynthesis calculations and the
variance of predictions from study to study are large enough that I usually
take these predictions as ``serving suggestions'' rather than recipes to
be followed to the letter.

In GCE, one thing my collaborators and I have tried to do is to separate
constraints on SSA and GSA by drawing on complementary observables.
Loosely speaking (see, e.g., Andrews et al. 2017, Weinberg et al. 2017), 
models predict that abundance ratio trends 
depend most strongly on yields (SSA),
while distribution functions depend on star formation history (SFH), 
stellar radial migration
(Schoenrich \& Binney 2009, Hayden et al. 2015, Loebman et al. 2016,
Johnson et al. 2021), and other GSA.

APOGEE abundances show that the median trends of [X/Mg] vs. [Mg/H] are
nearly universal throughout the Milky Way disk (Weinberg et al. 2019)
and bulge (Griffith et al. 2021a), provided one separately examines the
high-$\alpha$ and low-$\alpha$ populations.  This universality directly
demonstrates the dominant role of SSA in determining these trends, since
GSA features like SFH and gas flows are almost certainly
different between the outer disk, inner disk, and bulge.
Mg is a better reference element than Fe for this purpose because it has a 
single astrophysical source, massive stars exploding as core collapse
supernovae (CCSN), whereas Fe comes from both CCSN and Type Ia supernovae
(SNIa).  The separation between high-$\alpha$ and low-$\alpha$ populations
is really caused by the difference in Fe from SNIa, not by differences in 
$\alpha$ elements themselves.  For other elements, the separation of
the two [X/Mg] sequences depends on the relative contribution of 
CCSN (or, more generally, prompt sources) vs. SNIa (or, more generally,
delayed sources) to element X.  By formalizing this idea in a 
``2-process model,'' we have used observed trends in APOGEE and GALAH
to derive empirical constraints on the IMF-averaged yields of CCSN
and SNIa for many elements (Weinberg et al. 2019, 2022, Griffith et al. 2019,
2022).  For elements in which the delayed contribution comes from 
asymptotic giant branch (AGB) stars rather than SNIa, this decomposition
is only approximate.  Isolating CCSN yields from other contributions enables
sharper tests of theoretical models of CCSN and black hole formation
(Griffith et al. 2021b).

In a related vein,
Johnson et al. (2023) exploit the dependence of abundance
ratios on SSA to derive empirical constraints
on the metallicity dependence of nitrogen yields from the trend of
[N/O] vs. [O/H] observed in the Milky Way and external galaxies.
Separating the contributions of massive stars and intermediate mass stars
to the IMF-averaged yield is trickier, but we show that one can get 
useful empirical constraints on this decomposition from the separation 
of [N/O] ratios between high-$\alpha$ and low-$\alpha$ stellar populations
in the Milky Way disk (Vincenzo et al. 2021; J. Roberts et al., in prep.).

One key difference between present-day GCE and early '90s galaxy formation
is that abundance ``measurements'' are themselves derived by fitting
intricate stellar atmosphere models to observed spectra, making them
subject to complex theoretical and observational systematics.
Therefore, when deducing lessons about chemical evolution from abundance
data, one must also allow for the possibility of observational systematics
that are comparable in magnitude to GCE model differences.
One important example is the difference between [O/Fe] or [O/Mg] trends
found in optical vs. near-IR spectroscopic surveys
(Griffith et al. 2019, 2022), or even between analyses of the same
near-IR spectra that adopt different $T_{\rm eff}$ and log$g$ calibrations
(Hayes et al. 2022).  IMF-averaged CCSN yield models predict that [O/Mg]
is nearly independent of metallicity (Andrews et al. 2017), 
as found in APOGEE.  If the sloped
[O/Mg] trend found in optical surveys is correct, it has sharp implications
for the origin of O, Mg, or both.

\section{Yields, Outflows, and Equilibrium}
\label{sec:yields}

While abundance ratios provide a powerful tool for constraining yields,
or more generally SSA, with little sensitivity to galactic scale uncertainties,
there is a critical degeneracy between the overall scale of yields and the
mass-loading factor $\eta \equiv \dot{M}_{\rm out}/\mdotstar$ of galactic outflows.
This degeneracy can be easily understood in one-zone GCE models, which assume
a fully mixed gas reservoir so that the abundances of newly forming stars
depend on time but not on spatial position.  These models have a long
history (e.g., Talbott \& Arnett 1972, Larson 1972).  
Here I summarize some of the lessons from the analytic
solutions of Weinberg et al. (2017, hereafter WAF).

For an element like O (or Mg) whose production is dominated by massive
stars, one can approximate enrichment as instantaneous with a net
IMF-averaged yield $\yocc$ that includes CCSN and massive star winds.
The evolution of oxygen mass in the gas reservoir follows
\begin{eqnarray}
\Modot &=&  \yocc\mdotstar - \Zo\mdotstar - \eta\Zo\mdotstar + r\Zo\mdotstar \\
       &=&  \yocc\mdotstar - (1+\eta-r)\mdotstar\Zo~.
\label{eqn:modot}
\end{eqnarray}
Here $\Zo = M_{\rm O}/M_g$ is the oxygen mass fraction of the ISM, and the
terms on the r.h.s. represent new O production, loss of O from the ISM
to star formation and outflow, and return of O to the ISM by recycling
of stellar envelopes.  Although some recycling comes from lower mass
stars with long lifetimes, approximating it as instantaneous is usually
accurate compared to calculations that include full time dependence.
For a Kroupa (2001) IMF, the recycling factor $r \approx 0.4$, roughly
half of it from stars with $M > 8M_\odot$.

As first argued by Larson (1972), if there is ongoing gas accretion that
sustains star formation, then the ISM abundance does not grow indefinitely
but instead reaches an equilibrium when the source and sink terms of
Eq. 2
balance.  For an exponential star formation
history, $\mdotstar \propto e^{-t/\tausfh}$, and a time-independent
star formation {\it efficiency}, ${\rm SFE} = \taustar^{-1} = \mdotstar/M_g$,
one can show that the equilibrium abundance is
\begin{equation}
\Zoeq = {\yocc \over 1+\eta-r-\taustar/\tausfh}~.
\label{eqn:Zoeq}
\end{equation}
The full time-dependence is
\begin{equation}
\Zo(t) = \Zoeq \left(1-e^{-t/\bar{\tau}}\right)
\label{eqn:Zo}
\end{equation}
with
\begin{equation}
\bar{\tau} = {\taustar \over 1+\eta-r-\taustar/\tausfh}~.
\label{eqn:taubar}
\end{equation}
In the molecular gas of present-day star forming galaxies a typical SFE
timescale is $\taustar \approx 2\Gyr$ (Leroy et al. 2008, Sun et al. 2023),
so the correction $\taustar/\tausfh$ is often small compared to $1+\eta-r$.

Importantly, one can also find analytic solutions for Fe enrichment by
SNIa if one approximates the DTD as $R_{\rm Ia}(t) \propto e^{-(t-t_d)/\tauIa}$
after a minimum delay time $t_d$ (equation 53 of WAF), which allows one
to calculate the evolution of $\afe$ and $\feh$ for realistic cases.
The introduction of the new timescale $\tauIa$ makes the solutions more
complex than Eq. 4,
but the behavior remains intuitive.
One can accurately approximate a $t^{-1.1}$ power-law DTD (Maoz \& Graur 2017)
as a sum of two exponentials, and the presence of short-$\tauIa$ and 
long-$\tauIa$ components leads to interesting behavior in some scenarios.
WAF also present analytic solutions for a linear exponential SFH,
$\mdotstar \propto t e^{-t/\tausfh}$, and for cases with sudden changes
of $\eta$ or $\taustar$, and their equation (117) provides a general
way to model more complex star formation histories by combining the 
solutions for exponential histories.

For understanding the yield-outflow degeneracy, 
equations 3-5
convey the key points.
The ISM abundance in equilibrium depends on $\yocc$ {\it and} $\eta$
because the latter controls the rate at which metals are lost in galactic 
winds.  The timescale $\bar{\tau}$ for approaching equilibrium is typically short
compared to the age of the galaxy, though it can become long if $\eta$
is small and star formation is inefficient.

\begin{figure}[t]
  \includegraphics[height=2.5truein]{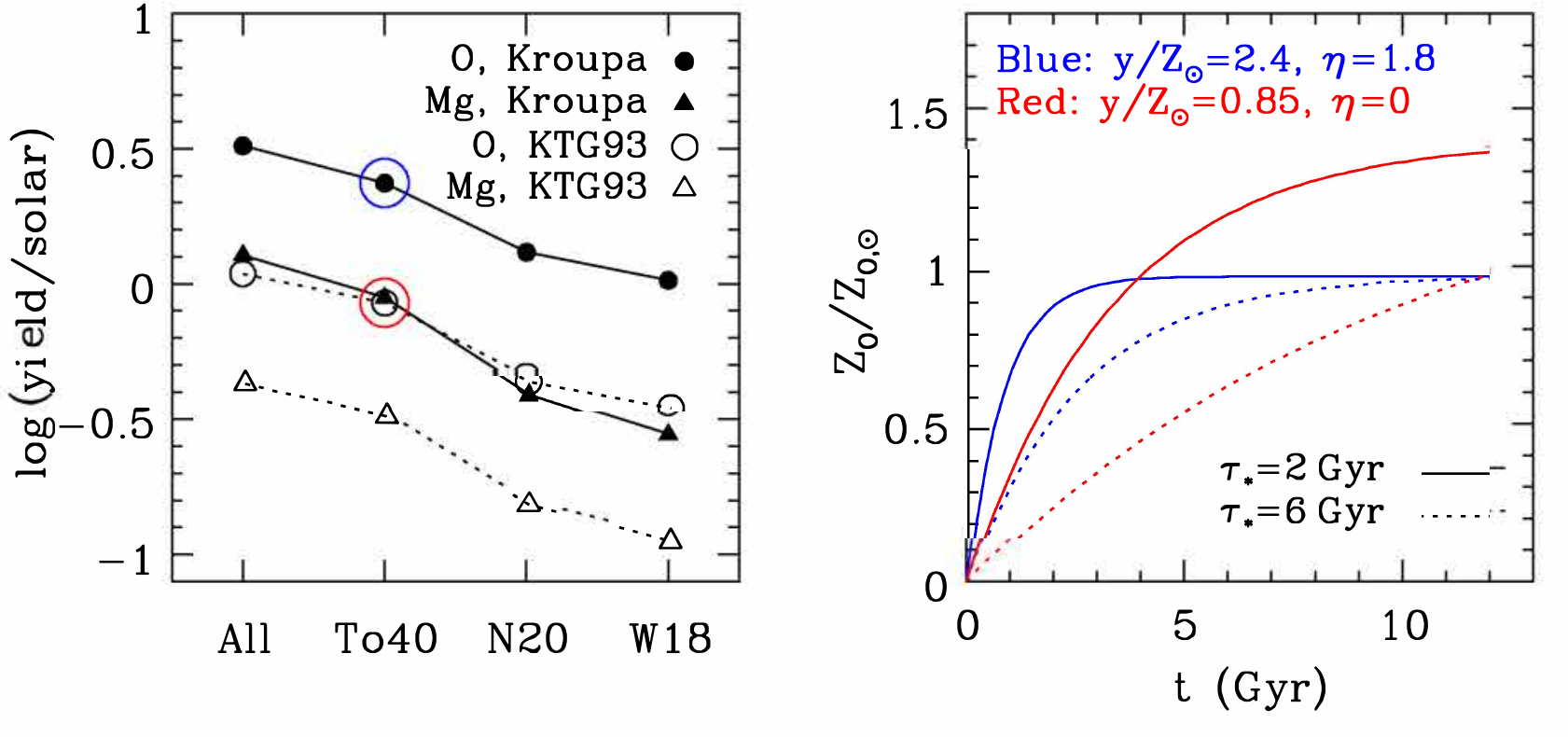}
  \caption{(Left) IMF-averaged net yield of O (circles) or Mg (triangles) computed from the CCSN models of Griffith et al. (2021b) for a Kroupa (2001) IMF (solid) or a Kroupa, Tout, \& Gilmore (1993) IMF (dotted), with four scenarios for black hole formation: all stars born with $8M_\odot \leq M \leq 120 M_\odot$ explode as CCSN; stars with $8M_\odot \leq M \leq 40 M_\odot$ explode; or the explosion landscapes predicted by Sukhbold et al. (2016) for two different neutrino-driven central engines.  Yields are scaled to the corresponding solar abundance.  For the same black hole scenario the predicted yields of the two IMFs differ by a factor of three.  For a given IMF, the difference between the scenarios with the most or least black hole formation is also a factor of three.  (Right) Evolution of the O abundance for the yields marked by the blue and red circle in the left-hand panel, $\yocc/Z_{{\rm O},\odot}=2.4$ and 0.85, respectively, with correspondingly chosen outflow mass loading factors $\eta=1.8$ or $\eta=0$.  Solid curves use an SFE timescale $\taustar=2\Gyr$ typical of molecular gas in local galaxies, while dotted curves use a longer SFE timescale $\taustar=6\Gyr$.  All curves use the analytic solution (Eq. 4)
  for an exponential SFH and assume $\tausfh \rightarrow \infty$ (constant SFR).}
  \label{fig:imf}
\end{figure}
For a Kroupa IMF and the CCSN yields of Chieffi \& Limongi (2004) and
Limongi \& Chieffi (2006), the predicted yield is $\yocc \approx 0.015$
(i.e., $1.5M_\odot$ of oxygen produced per $100M_\odot$ of stars formed).
Achieving solar oxygen abundance $Z_{{\rm O},\odot} \approx 0.006$
(Asplund et al. 2009) therefore requires strong outflows, 
with $\eta \approx 2$, a long-standing result in theoretical models
of the mass-metallicity relation (e.g., Finlator \& Dav\'e 2008,
Peeples \& Shankar 2011, Zahid et al. 2012) as well as my group's
GCE models (e.g., Andrews et al. 2017, WAF, Johnson et al. 2021).
I was initially puzzled that the ``Trieste group'' (which by now spans
many nations) was able to produce successful GCE models with no outflows
(e.g., Chiappini et al. 1997, Matteucci \& Francois 1998, Minchev et al. 2013,
Spitoni et al. 2019), but I eventually learned it is because
they adopt an IMF-averaged O yield that is about $3\times$ lower than mine,
with $\yocc \approx Z_{{\rm O},\odot}$ (Minchev, private communication).
In their case, the principal reason for this much lower yield is the
assumption of a steeper IMF (Miller \& Scalo 1979, Kroupa et al. 1993), with 
$\phi(M) \propto M^{-2.7}$ at high masses compared to $M^{-2.3}$ for a
Kroupa (2001) IMF.

In addition to the IMF, a key source of uncertainty in CCSN yields is
black hole formation (Sukhbold et al. 2016; Griffith et al. 2021b).
Production of O, Mg, and other $\alpha$-elements grows rapidly with
progenitor mass, but if the most massive stars implode to black holes
then they never release these products to the ISM.  The left panel of
Fig. 1,
adapted from Griffith et al. (2021b), shows their
IMF-averaged O and Mg yields for a Kroupa (2001) IMF or the much steeper
Kroupa et al. (1993) IMF, and for four different scenarios of black hole
formation: the complex ``landscapes'' predicted for two neutrino-driven
central engines (W18 and N20) by Sukhbold et al. (2016), a sharp 
transition from explosion to implosion at $M=40M_\odot$, or no black 
hole formation.  For a given IMF, the yields predicted for these four
scenarios span a factor of three, and for a given black hole scenario
the difference between these two IMFs is also a factor of three.

The right panel of Fig. 1
shows $\Zo(t)$ from 
Eq. 4
for two different combinations of yield and $\eta$:
$\yocc/\Zosun=2.4$ and $\eta=1.8$ (blue) and
$\yocc/\Zosun=0.85$ and $\eta=0$ (red).  In each case we assume
constant SFR ($\tausfh \rightarrow \infty$) and $\taustar=2\Gyr$ (solid)
or $6\Gyr$ (dotted).  As expected, both the high-yield/high-$\eta$ and
low-yield/low-$\eta$ combinations can produce $\Zo \approx \Zosun$ at
late times.  These examples and Eq. 5
also explain
why equilibrium is a central concept in my group's GCE models
(and earlier work such as Finlator \& Dav\'e 2008, Dav\'e et al. 2012)
but is not nearly so evident in the Trieste models.  If $\eta \approx 2$
then the timescale for reaching equilibrium is fairly short, even with
$\taustar=6\Gyr$ (blue dotted line).  If $\eta=0$ then the timescale is
longer, and $\Zo$ may still be climbing at $t=10-13\Gyr$.  For the
$\eta=0$, $\taustar=6\Gyr$ model (red dotted line), the abundance remains
well below equilibrium even at $t=12\Gyr$.
The impact of $\eta$ and $\taustar$ on time-evolution 
is amplified for Fe because of the delayed SNIa contribution.

The scale of yields is therefore tightly connected to the importance of
outflows {\it and} the emergence of equilibrium. Unfortunately, black hole
formation and other theoretical uncertainties make it difficult to predict
the absolute scale of CCSN yields with confidence.  Deuterium is one
possible way forward because its theoretical yield is perfectly known:
stars destroy whatever D they are born with, making $y_D=0$ (all deuterium
is primordial).  Unfortunately, the large sightline-to-sightline variance
of D/H (Linsky et al. 2006), plausibly interpreted as a sign of dust
depletion onto PAH molecules, makes determining the intrinsic ISM D/H
ratio difficult.  For the ISM D/H value advocated by Linsky et al. (2006),
the implied values of $\eta$ and $\yocc$ are fairly high, similar to the
blue models in Fig. 1
(Weinberg 2017).  The mean Fe yield
of CCSN (Rodriguez et al. 2022) offers another, empirical route towards
determining the overall yield scale (Weinberg, in prep.).  There are
uncertainties in going from the measured quantity to $\alpha$-element
yields such as $\yocc$, but the Rodriguez et al. measurement suggests a
value intermediate between the two cases shown in Fig. 1,
requiring mild outflows ($\eta \approx 1$) to reach solar abundance.

\section{Dimensionality and Scatter}
\label{sec:scatter}

Despite the yield-outflow degeneracy, abundance ratio trends
(e.g., [X/Mg] vs. [Mg/H]) provide strong empirical constraints on 
yield {\it ratios}, and by exploiting the separation of high-$\alpha$
and low-$\alpha$ populations one can start to separate the contribution
of distinct astrophysical processes (e.g., CCSN vs. SNIa vs. AGB).
The addition of stellar ages --- from asteroseismology, or parallax + 
photometry, or spectroscopic diagnostics trained on these methods ---
helps further separate prompt and delayed enrichment channels and
provides leverage on SFE and SFH.  Models with different SFH but the
same yields and SFE produce similar abundance ratio tracks, but they
produce different metallicity distribution functions because the number
of stars formed at a given metallicity is proportional to the SFR.
By fitting one-zone models to stellar abundances of dwarf spheroidals
or stellar streams one can get interesting SFH constraints from the
metallicity distribution alone (Sandford et al.\ 2022), and these become
sharper and more robust if the data include $\afe$ ratios and stellar ages
(Johnson et al. 2022).

With large, homogeneous data sets, one can measure the {\it scatter} around
mean abundance trends to gain additional diagnostics of yields, enrichment
history, mixing of stellar populations by mergers or migration, and mixing
of metals within the star-forming ISM.  Measuring intrinsic scatter accurately
is challenging because one must carefully assess the observational contribution,
including scatter caused by differential abundance systematics across the
stellar sample.  

Scatter in the high-dimensional space of stellar abundance ratios is a
nuanced concept, linked to the dimensionality of the stellar distribution 
in this space.  One of the earliest investigations of this dimensionality
was led by Yuan-Sen Ting, who showed that 6-9 principal components were
required to explain the abundance distribution of available high-resolution
data sets (Ting et al. 2012; for related approaches see Andrews et al. 2012,
2017 and Patil et al. 2022).  The dimensionality of the distribution is
connected to the number of distinct processes that contribute to the
observed abundances, but the connection is not a simple one.  For example,
in a one-zone model with a fully mixed gas reservoir, all stellar abundances
depend on a single parameter (time) even if many enrichment channels are
operating.  To produce scatter one needs multiple enrichment channels and
either imperfect ISM mixing (Krumholz \& Ting 2018)
or the mixing of stellar populations that
experience different relative contributions from these channels.

The success of the 2-process model in fitting individual abundance patterns
of Milky Way disk stars in APOGEE and GALAH shows that much of the
star-to-star variation is explained by two components, one describing overall
enrichment and a second describing the ratio of CCSN/SNIa enrichment
(Weinberg et al. 2022; Griffith et al. 2022).  Thus, to a surprisingly
good approximation one can predict all of a star's APOGEE or GALAH abundances
from its [Mg/H] and [Mg/Fe] alone.  In a similar fashion, one can predict
these abundances to surprising accuracy from [Fe/H] and age (Ness et al. 2019).
It would be interesting but ultimately disappointing if this 2-dimensional
description were perfect. 

Fortunately, it isn't.  My collaborators and I
have investigated this issue in two complementary ways, by examining the
conditional distribution $p({\rm [X/Fe]}|{\rm [Fe/H],[Mg/Fe]})$ and its
generalizations in APOGEE (Ting \& Weinberg 2022) and by examining the
residuals $\Delta([{\rm X/H}])$ of stellar abundances relative to a 2-process
fit in APOGEE and GALAH (Weinberg et al. 2022, Griffith et al. 2022).  
Each of these studies covers 15-16 elemental abundances.
Key findings include:

\begin{enumerate}
\item The observed total scatter (intrinsic + observational) around a 2-d
  fit for high-S/N APOGEE stars ranges from 
  $\approx 0.01-0.02$ dex (Mg, O, Si, Ca, Fe, Ni) to $\approx 0.1$ dex
  (Na, K, V, Ce), with a median of 0.03 dex.
\item While inferring intrinsic scatter from total scatter requires precise
  knowledge of observational uncertainties, the correlation of abundance
  residuals provides robust evidence for additional dimensions in 
  abundance space.
\item One must condition on at least seven elements (e.g., Fe, Mg, O, Ni,
  Si, Ca, Al) before the residual correlations of remaining APOGEE elements
  are consistent with observational noise.
\item Much of the $\afe$ scatter {\it within} the high-$\alpha$ and 
  low-$\alpha$ disk populations is caused by CCSN/SNIa variation within those
  populations at fixed [Fe/H]. These variations in turn are correlated with age.
\item APOGEE 2-process residuals show correlations among Ca, Na, Al, K, Cr, and Ce
  and separately among Ni, V, Mn, and Co.
\item GALAH 2-process residuals show correlations between Ba and Y, expected because
  both have large AGB $s$-process contributions, and significant correlations
  of these two with Zn.
\item Ba and Y in GALAH and Ce in APOGEE show increasing positive residuals
  at young stellar ages, implying a late-time enhancement of AGB enrichment
  relative to SNIa enrichment.  Intriguingly, Na shows a similar pattern
  in APOGEE even though it is not expected to have a large AGB contribution.
\item Residual abundances provide a powerful method for comparing stellar
  populations, including dwarf satellites, streams, and star clusters,
  because they control for star-by-star differences in overall metallicity
  and $\afe$ that can otherwise mask subtler differences in individual elements.
\end{enumerate}

\section{The Early History of the Milky Way}

Much of the discussion above focuses on the regime of the Milky Way
thin and thick disks.  The defining principle of IAU 377 was to connect
studies of the early Milky Way to those of high-redshift galaxies.  I will
conclude with three recent GCE studies of the early Milky Way, each of
which I mentioned in my talk.

While searches for low metallicity stars have often focused on the nearby
Galactic halo, theoretical models predict that the {\it oldest} stars
should be concentrated near the center of the Galaxy (White \& Springel 2000).
Deriving metallicities from \gaia\ BP/RP spectra with a machine learning
algorithm trained on APOGEE/\gaia\ overlap, Rix et al. (2022) show that
stars with $\monh < -1.5$ are indeed centrally concentrated, with a Gaussian
radial extent $\sigma_R$ of only 2.7 kpc.  Over the range 
$-2 \leq \monh \leq -1$ the observed metallicity distribution follows
$d\log n/d\monh = 1$.  In this regime, far below equilibrium, one can ignore
the loss terms in Eq.~2 and derive a fairly general prediction
\begin{equation}
    {d\log n \over d\monh} = \left[1-\left({M_*\over M_g}\right)
        \left({\taustar \dot{M}_g \over M_g}\right)\right]^{-1}
\label{eqn:mdf}
\end{equation}
The combination $(\taustar \dot{M}_g/M_g)$ corresponds to the 
fractional change of the gas reservoir mass over one SFE timescale.
The observed slope of unity matches the value theoretically expected for a
small stellar mass fraction and/or a slowly changing gas reservoir.

APOGEE disk abundances show a plateau at $\mgfe \approx 0.35$ for
stars with $-1.2 \leq \feh \leq -0.6$, which is conventionally
interpreted as representing the CCSN yield ratio.  However, after
using kinematic data to remove accreted halo stars, 
Conroy et al. (2022) find a more complex evolution in the H3 survey.
For $-2.5 \leq \feh \leq -1.5$ the typical $\afe$ declines from
$\approx 0.6$ to $\approx 0.3$, then rises slightly for 
$-1.5 \leq \feh \leq -0.5$.  With similar kinematic cuts, the APOGEE
data are consistent with this behavior.  Conroy et al. (2022)
interpret this trend with a model in which the true CCSN plateau
is at $\afe \approx 0.6$, the initial decline in $\afe$ is caused
by SNIa enrichment with a very low SFE ($\taustar \approx 50\Gyr$),
and the rise in $\afe$ for $\feh > -1.5$ is caused by rapidly 
accelerating SFE during which $\taustar$ drops from $50\Gyr$
to $2\Gyr$.  This ``simmer-to-boil'' transition also coincides with
a rapid increase in the angular momentum of stellar populations
over the range $-1.5 < \feh < -0.9$, also found in the analyses
of Belokurov \& Kravtsov (2022) and Rix et al. (2023).  While there
is still much to be decoded in these data, it appears that we are
seeing both the kinematic and chemical birth of the Milky Way disk
in these archaeological studies.

\begin{figure}[t]
  \includegraphics[height=2.5truein]{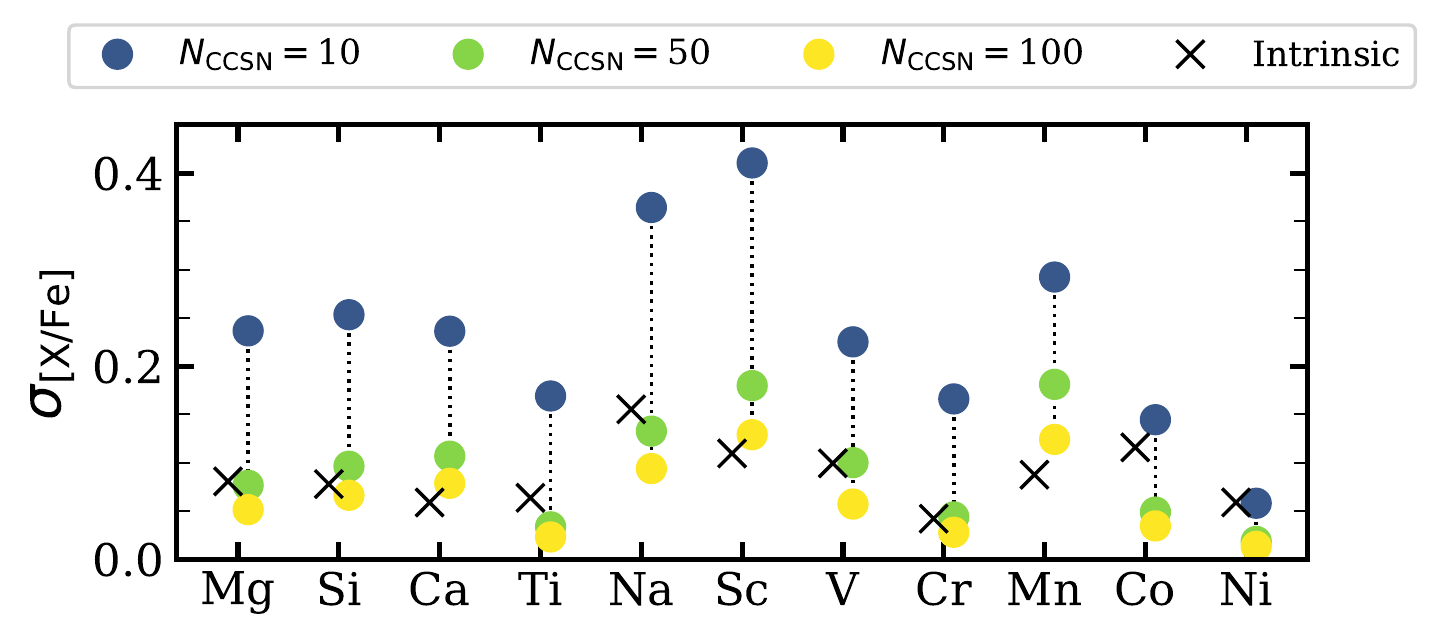}
  \caption{Crosses show the RMS intrinsic scatter in [X/Fe] at fixed [Fe/H] 
  measured in a sample of 86 subgiant halo stars with $-2 \leq \feh \leq -1$.
  Circles show the RMS scatter predicted for random draws from the high-mass
  IMF producing $N=10$ (blue), 50 (green), or 100 (yellow) CCSN, using the
  mass-dependent CCSN yields of Sukhbold et al. (2016).  For most elements,
  the measured scatter corresponds to the predicted scatter for $N\approx 50$.
  From Griffith et al. (2023).}
  \label{fig:scatter}
\end{figure}

The scatter of disk star abundances around 2-process predictions is small,
but we may reasonably expect scatter to be larger at low metallicities that
probe early phases of chemical evolution.  Griffith et al. (2023) measure
abundances of 12 elements in 86 metal-poor subgiants ($-2 \leq \feh \leq -1$),
with a survey strategy and analysis procedures designed specifically to
allow robust determination of the intrinsic scatter.  We find RMS intrinsic
scatter in [X/Fe] at fixed [Fe/H] ranging from 0.04 dex (Cr) to 0.16 dex (Na),
with a median of 0.08 dex.  One plausible interpretation of this scatter is
from stochastic sampling of the IMF, as any individual star is enriched 
not by the IMF-averaged CCSN yield but by the particular set of supernovae
that enriched its birth environment.

The Griffith et al. measurements are well explained if the number of CCSN
contributing to the enrichment of a given star is $N\sim 50$ (Fig.~2).
The stochastic CCSN model roughly (though not perfectly) predicts the
variation of scatter from element to element, and it also roughly explains
the correlations of element fluctuations from star to star (fig.~15 of
Griffith et al. 2023).  Producing the median metallicity of the sample
with $N=50$ CCSN requires that the products of these CCSN be diluted by
about $10^5 M_\odot$ of gas, orders of magnitude smaller than the likely
mass of the Galaxy at this epoch.  There are factor-of-three level 
uncertainties in this argument, and even the basic interpretation of scatter
as a stochastic IMF sampling effect may be incorrect.  Nonetheless, this
analysis illustrates what we can hope to learn by studying scatter
around abundance trends as well as the trends themselves.  If our stochastic
sampling interpretation is correct, then the early Galaxy was divided
into many ($10^4-10^6$) chemically disconnected regions, with each newly
formed star sampling only the nucleosynthetic products of its own
elemental horizon.

\section*{Acknowledgements}
\noindent 
I am grateful to the many collaborators from whom and with whom I have learned
the subject of GCE, especially (and in approximately chronological order)
Jennifer Johnson, Ralph Schoenrich, Brett Andrews, Jon Holtzman, 
Emily Griffith, James Johnson, Jon Bird, Fiorenzo Vincenzo, Yuan-Sen Ting,
Hans-Walter Rix, Juna Kollmeier, Charlie Conroy, Nathan Sandford,
Dan Weisz, and many others in the SDSS collaboration.
I thank the IAU organizers for the opportunity to present some of this work
to an amazing group of participants from around the world.
My GCE work has been supported by the U.S. National Science Foundation,
most recently by NSF AST-1909841.  It has relied critically on data from
the Sloan Digital Sky Survey ({\tt www.sdss.org}).

\end{document}